\documentclass[two column,showpacs]{revtex4}
\usepackage{graphicx}
\usepackage{dcolumn}

\begin{document}


\title{Contour integral method for thermal and quantal fluctuations}

\author{K. Kaneko$^{1}$ and M. Hasegawa$^{2}$}
\affiliation{
$^{1}$Department of Physics, Kyushu Sangyo University, Fukuoka 813-8503, Japan \\
$^{2}$Laboratory of Physics, Fukuoka Dental College, Fukuoka 814-0193, Japan 
}

\date{\today}

\begin{abstract}
The partition function by means of the static path approximation (SPA) plus 
the random-phase approximation (RPA) treatment can be written as a contour integral form 
without solving the RPA equations for a separable interaction. This method is an efficient 
way to evaluate numerically the partition function for realistic calculations. 
As an illustration, we adopt the pairing model at finite temperature. 

\end{abstract}


\maketitle

Mean-field approximations, such as the finite temperature Hartree-Fock Bogolyubov (FTHFB) 
approach \cite{Sano,Goodman,Egido}, 
are standard methods for describing nuclei at finite temperature, and the collective 
motions around the static HFB solution are obtained by solving the finite temperature 
random phase approximation (RPA). 
At zero temperature, the HF plus RPA method gives a good approximation to 
ground-state energy of the exact shell model calculation \cite{Stetcu}. 
However, the FTHFB+RPA method is not applicable to systems at high temperature 
because of large thermal fluctuations. 
The inclusion of fluctuations and correlations beyond thermal 
mean field can be done by use of the path integral representation of the partition function. 
The static path approximation (SPA) \cite{Alhassid,Bertsch,Rossignoli1}
is useful treatment to evaluate approximately the 
partition function in finite systems with separable interactions, and provides the exact 
result at high temperature. However, as temperature decreases the SPA becomes inaccurate 
because the quantal fluctuations around the mean fields can no longer be neglected. 
Small amplitude fluctuations around such a configuration 
give corrections to the partition function similar to the standard RPA around the mean field. 
The SPA+RPA method \cite{Puddu, Lauritzen,Rossignoli,Attias} can take the effects into 
account. 
It should, however, be pointed out that in the limit of zero temperature the formalism 
breaks down since it includes contributions of small-amplitude quantal fluctuations around 
unstable mean-fields. 

In realistic calculations for deformed nuclei or heavy nuclei, the evaluation of 
the RPA partition function is difficult due to 
time-consuming computation, because a complete set of RPA frequencies and amplitudes must 
be known with high accuracy for each temperature and the number of the RPA eigenvalues 
is very large. Recently, a contour integral representation for the standard RPA correlation 
energy \cite{Donau} and more efficient way \cite{Shimizu} based on the linear response theory 
have been proposed. 
These methods are valuable and useful for evaluating the standard RPA correlation energy. 
In this letter, we present a contour integral method for the RPA partition 
function at finite temperature. 

We consider the total Hamiltonian
\begin{eqnarray}
H & = & H_{0} + V, 
\label{eq:1}
\end{eqnarray}
where $H_{0}$ and $V$ are the single-particle Hamiltonian and a two-body residual 
interaction, respectively. Here, the two-body interaction is assumed to be of separable 
multipole-multipole form 
interaction: 
\begin{eqnarray}
V & = & -\frac{1}{2}\sum_{\alpha}\chi_{\alpha}Q_{\alpha}^{2},
\label{eq:2}
\end{eqnarray}
where $Q_{\alpha}$ is a one-body Hermitian operator 
\begin{eqnarray}
Q_{\alpha} & = & \sum_{kl}q_{kl}^{\alpha}c_{k}^{\dag}c_{l}. 
\label{eq:3}
\end{eqnarray}
Using the Hubbard-Stratonovich transformation \cite{Hubbard}, the functional integral 
representation for the partition function is written as an auxiliary field path-integral
\begin{eqnarray}
Z & = & {\rm Tr}(e^{-\beta H}) \nonumber \\
  & = & \int {\cal D}[\sigma_{\alpha}(\tau)]{\rm exp}\left( -\frac{1}{2}\sum_{\alpha}\chi_{\alpha}
  \int_{0}^{\beta}\sigma_{\alpha}^{2}(\tau)d\tau \right) \nonumber \\
  &   & \times {\rm Tr}[\hat{T} {\rm exp}(-\int_{0}^{\beta}d\tau H'(\tau)], \\
H'(\tau) & = &  H_{0} - \sum_{\alpha}\chi_{\alpha}\sigma_{\alpha}(\tau)Q_{\alpha},
\label{eq:4}
\end{eqnarray}
where $\beta=1/T$ is the inverse temperature, and $\hat{T}$ denotes time ordering operator. 
Let us now use the Fourier transformation of the auxiliary fields $\sigma_{\alpha}$ 
defined as 
\begin{eqnarray}
\sigma_{\alpha}(\tau) & = & \bar{\sigma}_{\alpha} + \sum_{n\neq 0}\eta_{\alpha n}
e^{-i\omega_{n}\tau}, 
\label{eq:5}
\end{eqnarray}
with the Matsubara frequencies $\omega_{n}=2\pi n/\beta$ and 
the static variables $\bar{\sigma}_{\alpha}$. The SPA is obtained 
by considering only the static paths in the evolution operator. 
The Hamiltonian $H'(\tau)$ can be divided into the static and fluctuation parts: 
\begin{eqnarray}
H'(\tau) & = & H_{s} + \delta H(\tau),
\label{eq:6}
\end{eqnarray}
where $H_{s}$ and $\delta H(\tau)$ are the static and fluctuation part as follows:
\begin{eqnarray}
H_{s} & = & H_{0} - \sum_{\alpha}\chi_{\alpha}\bar{\sigma}_{\alpha}Q_{\alpha}, \\
\delta H(\tau) & = & - \sum_{\alpha}\chi_{\alpha}\sum_{n\neq 0}\eta_{\alpha n}Q_{\alpha}{\rm e}^{-i\omega_{n}\tau}. 
\label{eq:7}
\end{eqnarray}
Substituting Eqs. (6-9) into Eq. (4) and performing the Gauss 
integral over the coefficients $\eta_{\alpha n}$, we obtain 
\begin{eqnarray}
Z & = & \frac{\beta\chi_{\alpha}}{2\pi} \int d\bar{\sigma}_{\alpha} 
{\rm exp}\left( -\frac{1}{2}\sum_{\alpha}\chi_{\alpha}\bar{\sigma}_{\alpha}^{2} \right) \nonumber \\
& & \times\prod_{k}(1+e^{-\beta\varepsilon_{k}})C_{RPA}, \\
C_{RPA} & = & \prod_{n > 0} {\rm det}(1 - \chi R(\omega_{n}))^{-1},
\label{eq:8}
\end{eqnarray}
where $\varepsilon_{k}$ are the eigenenergies of $H'$, and $R(\omega_{n})$ is the response 
function matrix given as 
\begin{eqnarray}
R_{\alpha\alpha'}(\omega_{n}) & = & \sum_{kl}\frac{q_{kl}^{\alpha}q_{kl}^{\alpha'}\varepsilon_{kl}(f_{k}-f_{l})}
{\varepsilon_{kl}^{2}+\omega_{n}^{2}},
\label{eq:9}
\end{eqnarray}
with $\varepsilon_{kl}=\varepsilon_{k}-\varepsilon_{l}$ and the Fermi occupation probability 
$f_{k}=(1+e^{\beta \varepsilon_{k}})^{-1}$. 
Solving the dispersion equation 
${\rm det}(1 - \chi R(i\omega_{n}))=0$ or diagonalizing the finite temperature RPA 
equations, $C_{RPA}$ can be written exactly in terms of the RPA frequencies 
$\Omega_{\nu}$, 
\begin{eqnarray}
C_{RPA} & = & \frac{\prod_{kl}'\frac{1}{\varepsilon_{kl}}{\rm sinh}\frac{\beta\varepsilon_{kl}}{2}}
{\prod_{\nu}\frac{1}{\Omega_{\nu}}{\rm sinh}\frac{\beta\Omega_{\nu}}{2}} ,
\label{eq:10}
\end{eqnarray}
where the prime in $\prod_{kl}'$ restricts the product to pairs $(k,l)$ that satisfy 
$k < l$ and $\varepsilon_{kl}\neq 0$. 
Note that for deformed nuclei or heavy nuclei there are many RPA frequencies 
$\Omega_{\nu}$ in the numerator and $\varepsilon_{kl}$ in the denominator. 
For arbitrary $\sigma$, the lowest $\Omega_{\nu}$'s may become zero and imaginary (or complex) 
at low temperature. However, $C_{RPA}(\sigma)$ remains finite and positive both for 
$\Omega_{\nu}=0$ and for imaginary $\Omega_{\nu}$ if $\beta |\Omega|<2\pi$. This condition 
will be violated at very low temperature. 

Let us define the thermal RPA energy 
\begin{eqnarray}
E_{RPA} & = &  -\frac{\partial {\rm ln}C_{RPA}}{\partial \beta}  \nonumber \\
        & = &  \frac{1}{2}\left( \sum_{\nu}\Omega_{\nu}{\rm coth}\frac{\beta\Omega_{\nu}}{2} 
        -\sum_{k < l}\varepsilon_{kl}{\rm coth}\frac{\beta\varepsilon_{kl}}{2}\right), \nonumber \\ 
\label{eq:11}
\end{eqnarray}
where we neglected the temperature dependence of the RPA frequencies $\Omega_{\nu}$. 
In the limit $\beta \rightarrow \infty$, the thermal RPA energy $E_{RPA}$ reduces correctly 
the RPA correlation energy 
\begin{eqnarray}
E_{RPA}(T=0)  & = &  \frac{1}{2}\left( \sum_{\nu}\Omega_{\nu} - \sum_{k < l}\varepsilon_{kl}\right).
\label{eq:12}
\end{eqnarray}

\begin{figure}[t]
\includegraphics[width=8cm,height=10cm]{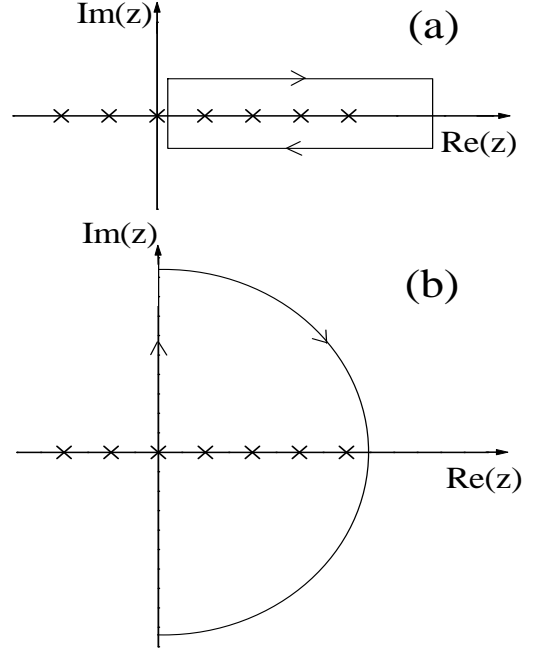}
  \caption{Contours used in evaluating the integrals (19) and (20). 
  Crosses denote the positions of all the RPA roots $\Omega_{\mu}$ and the poles 
  $\varepsilon_{kl}$}
  \label{fig1}
\end{figure}

We now introduce the analytical function of the complex vaiable $z$
\begin{eqnarray}
f(z) & = & {\rm det}(1 - \chi R(iz))=\frac{\prod_{\nu}(\Omega_{\nu}^{2}-z^{2})}
{\prod_{kl}'(\varepsilon_{kl}^{2}-z^{2})}. 
\label{eq:13}
\end{eqnarray}
As mentioned above, the FTRPA frequencies $\Omega_{\nu}$ are obtained as the zeros of 
the determinant, ${\rm det}(1 - \chi R(i\omega))=0$, {i.e.,} $f(\omega)=0$. 
Here, we notice that $\omega=\pm\Omega_{\nu}$ are the roots of $f(\omega)=0$ 
and the function $f(\omega)$ has obviously poles at $\omega=\pm\varepsilon_{kl}$. 
Thus we obtain the following relation
\begin{eqnarray}
\frac{f(z)'}{f(z)} & = & \frac{d}{dz}{\rm log}f(z) \nonumber \\
  & = & \sum_{k < l}\left( \frac{1}{\varepsilon_{kl}-z}-\frac{1}{\varepsilon_{kl}+z}\right) \nonumber \\
  & - & \sum_{\nu}\left( \frac{1}{\Omega_{\nu}-z}-\frac{1}{\Omega_{\nu}+z}\right), 
\label{eq:14}
\end{eqnarray}
with a derivative $f(z)'=df(z)/dz$. 
Applying Cauchy's theorem to a meromorphic function $f(z)$, we obtain the following relation
\begin{eqnarray}
\frac{1}{2\pi i}\oint dz g(z)\frac{f(z)'}{f(z)} & = & \sum_{k < l}g(\varepsilon_{kl}) - 
\sum_{\nu}g(\Omega_{\nu}),
\label{eq:15}
\end{eqnarray}
where $g(z)$ is an arbitrary complex function which is analytical in the enclosed region. 
The contour of the integration goes around all positive roots and all poles of $f(z)$ 
as shown in Figs. 1(a) and (b), where we consider the case of real RPA frequencies. 
The thermal RPA energy $E_{RPA}$, Eq. (14), is written as the closed contour integral
\begin{eqnarray}
E_{RPA} & = & \frac{-1}{4\pi i}\oint dz (z{\rm coth}\frac{\beta z}{2})\frac{f(z)'}{f(z)}, \\
 & = & \frac{1}{4\pi i}\oint dz (z{\rm coth}\frac{\beta z}{2})'{\rm ln}f(z), 
\label{eq:16}
\end{eqnarray}
where the closed path for Eq. (19) is taken so that the function 
$z{\rm coth}\beta z/2$ is analytical in the closed region(see Fig. 1(a)). 
A modified contour integral for Eq. (20) is illustrated in Fig. 1(b), 
where the contribution from the semicircle vanishes in the limit of infinite radius. 
These are the extended form of the contour integral of the RPA correlation 
energy at zero temperature \cite{Donau, Shimizu}. 
Integrating 
$E_{RPA} =  -\partial {\rm ln}C_{RPA}/\partial \beta$  over $\beta$ 
in Eq. (14) provides the RPA partition function $C_{RPA}$ 
\begin{eqnarray}
C_{RPA} & = & {\rm exp}\left( -\int_{0}^{\beta}E_{RPA}d\beta'\right). 
\label{eq:17}
\end{eqnarray}
In comparison with the usual RPA method of Eq. (14), 
the advantage of the contour integral form in Eqs. (19-20) is that 
one can choose the integration path so that the integrand is smooth enough and the 
mesh of numerical integration along such path can be much larger than the actual energy 
intervals of the FTRPA solutions $\Omega_{\nu}$. The contour integral calculation reduces 
drastically the computation time without loss of precision. 
In practical calculations of the above method, we evaluated the contour integral (20) 
illustrated in Fig. 1(b) neglecting the function $(z{\rm coth}\beta z/2)'$. 
The calculated results in Figs. 2-4 agree well with those of the SPA plus RPA calculation 
with Eq. (14). However, in general the neglect of the factor $(z{\rm coth}\beta z/2)'$ 
is not justified in all cases. In that case, this factor should be suitably taken into 
account in Eqs. (19) and (20). 

As an illustration, we consider a monopole pairing Hamiltonian 
\begin{eqnarray}
H & = & \sum_{k}\varepsilon_{k}(c_{k}^{\dag}c_{k}+c_{\bar{k}}^{\dag}c_{\bar{k}})-GP^{\dag}P, 
\label{eq:17}
\end{eqnarray}
with $\bar{k}$ the time reversed states.
Here, $\varepsilon_{k}$ is a single-particle energy and $P$ is 
the pairing operator $P=\sum_{k}c_{\bar{k}}c_{k}$. 
By means of the SPA+RPA \cite{Rossignoli,Attias} based on the Hubbard-Stratonovich 
transformation \cite{Hubbard}, the canonical partition function is given by
\begin{eqnarray}
Z_{N}^{\rm c} & = & {\rm Tr}[P_{N}e^{-H/T}]_{\rm SPA+RPA} \nonumber \\
& = & \frac{2}{GT}\int_{0}^{\infty}\Delta d\Delta 
e^{-\Delta^{2}/GT}Z_{N}C_{\rm RPA}. 
\label{eq:18}
\end{eqnarray}
\begin{eqnarray}
Z_{N} & = & \frac{1}{2}\prod_{k}e^{-\gamma_{k}/T}(1+e^{-\lambda_{k}/T})^{2} \nonumber \\
& & \times[1+\sigma \prod_{k'}{\rm tanh}^{2}(\lambda_{k'}/T)], \\
C_{\rm RPA} & = & \prod_{k}\frac{\omega_{k}{\rm sinh}[\lambda_{k}/T]}
{2\lambda_{k}{\rm sinh}[\omega_{k}/2T]}, 
\label{eq:19}
\end{eqnarray}
where we introduced the number parity projection $P_{N}=(1+\sigma e^{i\pi N})/2$ 
($\sigma$ means the even or odd number parity)\cite{Rossignoli} instead of the exact 
number projection. 
Here use the notation 
$\omega_{k}$ for the conventional thermal RPA energies and 
$\lambda_{k}=\sqrt{\varepsilon'^{2}_{k}+\Delta^{2}}$, 
$\varepsilon'_{k}=\varepsilon_{k} - \mu - G/2$, and 
$\gamma_{k}=\varepsilon_{k} - \mu - \lambda_{k}$. 
The SPA partition function is obtained by neglecting the RPA partition function $C_{RPA}$. 
Then the thermal energy can be calculated from 
$E=-\partial {\rm ln}Z_{N}^{\rm c}/\partial\beta$. 
In this calculation, we use the single-particle energies $\varepsilon_{k}$ given by 
an axially deformed Woods-Saxon potential with spin-orbit interaction \cite{Cwoik}. 
The Woods-Saxon parameters are chosen so as to approximately reproduce 
the experimental single-particle 
energies extracted from the energy levels of the odd nucleus $^{133}$Sn 
($^{132}$Sn core plus one neutron). 
The deformation parameter estimated from the experimental 
$B(E2)$ value is $\delta=0.23$ in the even-even nucleus $^{184}$W. 
The 50 doubly degenerate single-particle levels are taken outside 
$^{132}$Sn core, and we fix the pairing force strength at $G=20/A$ MeV. 
As mentioned above, the SPA+RPA breaks down at low temperature. 
However, it has recently been shown that in the monopole pairing case the SPA+RPA with 
the number parity projection reproduces well exact results for low temperature 
\cite{Rossignoli}. Thus, the number parity projection is essential to describe the thermal 
properties for low temperature. 

\begin{figure}[t]
\includegraphics[width=8cm,height=10cm]{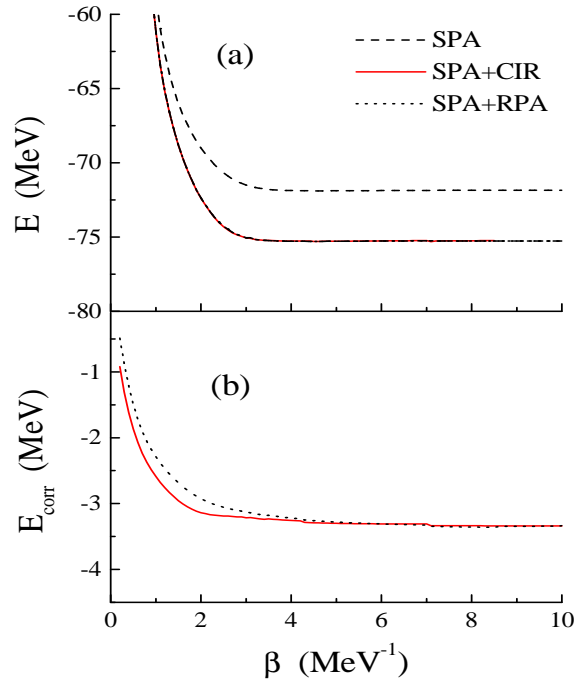}
  \caption{(Color online) Thermal energy and thermal correlation energy as a function $\beta$. 
          The dashed line denotes the SPA, the solid line the SPA+CIR, and the 
          broken line the SPA+RPA. }
  \label{fig2}
\end{figure}
\begin{figure}[t]
\includegraphics[width=8cm,height=10cm]{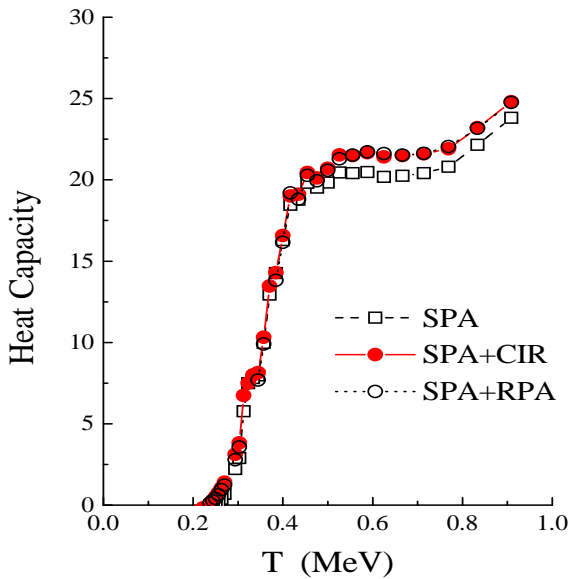}
  \caption{(Color online) Heat capacity as a function of $\beta$. 
  The open squares denote the SPA, the solid circles the SPA+CIR, and the 
  open circles the SPA+RPA.}
  \label{fig3}
\end{figure}
\begin{figure}[t]
\includegraphics[width=8cm,height=10cm]{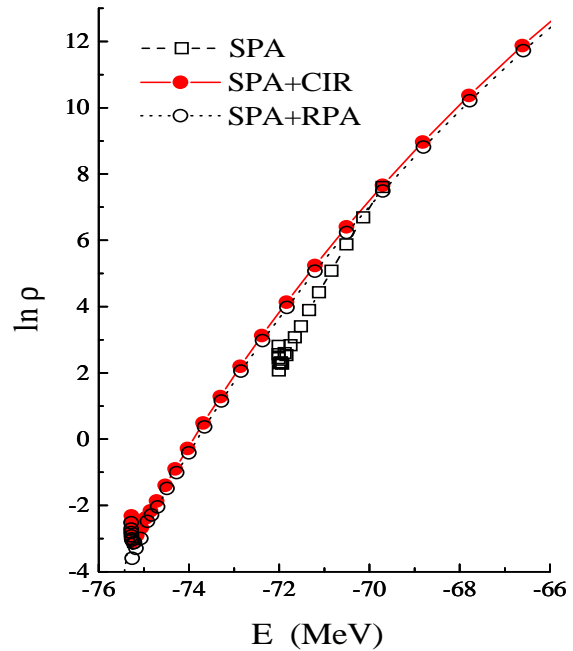}
  \caption{(Color online) Level density as a function of thermal energy $E$. 
  The same symbols as those in Fig. 3 are used.}
  \label{fig4}
\end{figure}

For the partition function (23), we performed the three cases of numerical calculations; 
the SPA, the SPA plus contour integral representation 
(SPA+CIR) with Eq. (19-20), and the SPA plus RPA (SPA+RPA) with Eq. (14) using the RPA frequencies. 
The thermal energies are calculated from $E=-\partial {\rm ln}Z_{N}^{\rm c}/\partial\beta$. 
The thermal energy obtained in the SPA+CIR (SPA+RPA) deviates from that of the SPA. 
The energy difference between the SPA+CIR (SPA+RPA) and the SPA is 
the correlation energy $E_{corr}= E - E_{SPA}$. 
Figure 2 shows the thermal energy $E$ and the RPA correlation energy $E_{corr}$ as 
a function of $\beta$. 
We can see that the SPA+CIR results agree well with the SPA+RPA one. 
Thus, the contour integral method is useful for studying the thermal and quantal fluctuations. 
For comparison, we solved the Richardson 
equations \cite{Richardson,Hasegawa} and calculated exactly the ground-state energy $E_{gs}$. 
The obtained energy$E_{gs}=-74.48$ MeV is quite close to the SPA+CIR value, $E=-75.24$ MeV 
at $\beta=10$ ${\rm MeV^{-1}}$, supporting the SPA+CIR method. 

We can see that the heat capacity exhibits the characteristic S-shape behavior as shown in 
Fig. 3. This S-shape heat capacity was recently observed \cite{Schiller,Melby} 
in $^{162}$Dy, $^{166}$Er, 
and $^{172}$Yb, and was interpreted as an signature of the breaking of nucleon Cooper pairs. 
The SPA+RPA and SPA+CIR results deviate from the SPA one in higher 
temperature region above $T=$0.5 MeV, and the quantal fluctuations are important 
for this region. 
In our previous paper \cite{Kaneko,Kaneko1}, we clarified that 
the S-shape behavior of heat capacity is attributed to the reduction of the pairing energy 
which can be calculated from 
$G\langle P^{\dag}P \rangle = GT\partial {\rm ln}Z_{N}^{\rm c}/\partial G$. 

When problems are associated with the variable $N$, they can be solved by considering 
a canonical ensemble, in which the particle number is strictly fixed. 
Since we used the number parity projection instead of the exact number projection, 
the level density for a system with $N$ particles is given by an inverse Laplace 
transformation of the partition function in Eq. (23)
\begin{eqnarray}
\rho(E,N) & = & \frac{1}{2\pi i}\int_{\beta_{0}-i\infty}^{\beta_{0}+i\infty}d\beta 
e^{\beta E}Z^{\rm c}_{N}(\beta). 
\label{eq:20}
\end{eqnarray}
In the saddle point approximation for the $\beta$ integral, the level density is given by 
\begin{eqnarray}
\rho(E,N) & \approx & \frac{Z^{\rm c}_{N}e^{\beta E}}{[2\pi
\partial^{2}{\rm ln}Z^{\rm c}_{N}/\partial\beta^{2}]^{1/2}}. 
\label{eq:21}
\end{eqnarray}
Figure \ref{fig4} depicts the natural logarithm of the level density (\ref{eq:21}) 
as a function of the thermal energy. 
We can see that the SPA+CIR result agrees with the SPA+RPA one. 

In conclusion, we developed the contour integral method which drastically reduces the 
computation time without loss of precision when evaluating the thermal and quantal fluctuations. 
This method is the extension of contour integral representation at zero temperature to 
the thermal system, and is numerically efficient. In this paper, we considered the simple 
pairing model as an illustration. The contour integral method is also applicable to 
more realistic interactions such as the pairing plus quadrupole-quadrupole interaction. 
This study is now in progress.



\end{document}